\documentclass{article}
\usepackage[cp1251]{inputenc}
\usepackage[dvips]{graphics}\usepackage{array,hhline,amssymb,amsthm}
\def\L{{\mathfrak{L}}}
\def\O{{\mathcal{O}}}  \def\o{{\mathbf{o}}}
\def\U{\mathfrak{O}}   \def\u{\mathfrak{o}}
\def\R{\mathbb{R}}

\def\tr{{\rm tr}} 
\def\buu#1{\buildrel{*}\over{#1}}

\def\bux#1#2{\buildrel{*}\over{#1}_{#2}}

\title{Generalization of Lagrangian of electro-weak interaction
to the octonionic algebra}

\author{V. Yu. Dorofeev\thanks{Dep. of Math., SPb SUEF,
Sadovaya 21, 191023, St.Petersburg,
Russia, E-mail: friedlab@mail.ru}\\
Theoretical Physics laboratory, devoted to Friedmann}
\begin{document}
\maketitle
\begin{abstract}
In the article it is considered the extension of Weinberg-Salam
theory from SU(2) group to the octonionic algebra. The extended
octonionic algebra is used as particle wave function instead of
spinors on su(2). It is shown, that leads to appearance of a new
neutral massive vector C and E-bosons (peculiar property of the
E-boson is that it does not interact with matter), and two charged
massive vector $D$ and $D^*$ bosons.
\end{abstract}
\section*{Introduction}
The progress in extending physical theory to octonionic algebra is
related with developing of works \cite{Zorn} and \cite{Daboul},
where useful matrix representation of non-associative algebra was
suggested. The tax for using matrix view was an unusual
multiplication law. Therefore, according to Feinmann interpretation
of quantum particle motion in extern field, although there is no
charge, during the particle movement its condition alternates and
the alternation is caused by non-associative interaction nature.

Perhaps, the first time octonionic algebra was applied to the
question in work \cite{Jordan}, where octonionic the science of
quantum mechanics was introduced.

Some scientists connect extended algebra with extended space-time
interpretation \cite{Nesterov}. Also, a range of scientists regard
the extension as deriving new properties in the network of
electromagnetic theory (e.g. in \cite{Gobre}).

The interpretation of this new type of interaction, in terms of
fields, could be found in \cite{Fradkin}. Meanwhile, the research
over the relation of the new types of interaction with the
conventional ones, as far as the author is concerned, has not ever
taken place.

\section{Octonionic algebra}
The doubling of quaternionic algebra leads, in particular, to
octonionic  algebra -- $\O$, which is a linear space over the field
of real numbers $\R$, for any $\o$ from octonionic algebra $\O$ has
a linear representation
$$\o=\sum_{j=0}^7\alpha^je^j,\quad \alpha_j\in\R,
j=0,1,2,\dots,7.$$

The multiplication is defined through generatrices $e^k\in\O$. In
particular,
$$(e^0)^2=1,\quad(e^j)^2=-1,\quad j=1,2,\dots,7,$$
so the first component is a real number while the others are to be
considered as complex units.

Provided we denote $\hat e^k=e^{k+4}, k=1,2,3$, the multiplication
on octonionic algebra is defined in the following way \cite{Daboul}
($i,j,k=1,2,3$):
$$e^ie^j=-\delta^{ij}e^0+\varepsilon^{ijk}e^k$$
$$\hat e^i\hat e^j=-\delta^{ij}e^0-\varepsilon^{ijk}e^k$$
\begin{equation}\label{bao}
e^i\hat e^j=-\delta^{ij}e^4-\varepsilon^{ijk}\hat e^k
\end{equation}
$$e^ie^4=\hat e^i,\quad e^4\hat e^i=e^i,$$
where the entirely antisymmetrical about its indexes permutation
symbol is introduced $\varepsilon^{123}=1$. It is easy to ensure,
that the multiplication law for generatrices leads to
non-associative algebra. Also,
\begin{equation}\label{asst}
\{e^i,e^j,e^k\}=(e^ie^j)e^k-
e^i(e^je^k)=2\varepsilon^{ijkl}e^l,\quad i,j,k,l=1,2,\dots,7,
\end{equation}
where $\varepsilon^{ijkl}$ is entirely antisymmetrical symbol, which
is equal to the unit for the following expressions:
\begin{equation}\label{ind}
1247,\qquad1265,\qquad2345,\qquad2376,\qquad3146,\qquad3157,\qquad4567.
\end{equation}

Octonionic algebra cannot be represented by matrices with
traditional multiplication rule, but the special multiplication rule
can be introduced, which admits such representation. In the capacity
of these matrices one unit matrix corresponding to the unit element
$e^0$ can be chosen along with seven matrices $\tilde\Sigma_k$ for
which the multiplication law $*$ is introduced as follows
($i,j,k=1,2,\dots,7$):
\begin{equation}\label{nonas1}
\tilde\Sigma^i*\tilde\Sigma^j=-\delta^{ij}+\varepsilon^{ijk}\tilde\Sigma^k,
\end{equation}
where entirely antisymmetrical symbol $\varepsilon^{ijk}$ is not
null if only
\begin{equation}\label{nonasalg}
\varepsilon^{123}=\varepsilon^{145}=\varepsilon^{176}=\varepsilon^{246}=
\varepsilon^{257}=\varepsilon^{347}=\varepsilon^{365}=1.
\end{equation}

Instead of matrices $\tilde\Sigma^k,k=1,2,\dots,7$ it is more
convenient to use $\Sigma^k=i\tilde\Sigma^k,k=1,2,\dots,7$, for
which, as it follows from (\ref{nonas1}), multiplication can be
defined as follows:
\begin{equation}\label{nonas2}
\Sigma^i*\Sigma^j=\delta^{ij}+i\varepsilon^{ijk}\Sigma^k, \quad
i,j,k=1,2,\dots,7.
\end{equation}

Introduce the matrices $\Sigma^i$ as it is in \cite{Zorn} $i=1,2,3$:
\begin{equation}\label{Zornpr}
\matrix{\Sigma^0=\left(\matrix{1&0\cr0&1}\right)&\Sigma^i=
\left(\matrix{0&-i\sigma^i\cr i\sigma^i&0}\right)\cr
\Sigma^4=\left(\matrix{-1&0\cr0&1}\right)&
\Sigma^{4+i}=\left(\matrix{0&-\sigma^i\cr -\sigma^i&0}\right)},
\end{equation}
where $\sigma^i,i=1,2,3$ are Pauli matrices:
\begin{equation}\label{matDir}
\sigma^1=\left(\matrix{0&1\cr1&0}\right),\qquad
\sigma^2=\left(\matrix{0&-i\cr i&0}\right),\qquad
\sigma^3=\left(\matrix{1&0\cr0&-1}\right).
\end{equation}

Along with $\Sigma^i$ matrices lets introduce abstract matrices
$\U\supset \O$ according to the rule, claiming any abstract matrix
$\u\in\O$ looks like:
\begin{equation}\label{om1}
\u=\left(\matrix{\lambda I&A\cr B&\xi I}\right),
\end{equation}
where $A,B$ are complex matrices, $(2\times2)$, $\lambda,\xi$ are
complex functions and $I$ is a unit matrix $(2\times2)$.

The abstract matrices are summed up as follows:
\begin{equation}\label{Dab}
\u+\u'=\left(\matrix{\lambda I&A\cr B&\xi I}\right)+
\left(\matrix{\lambda' I&A'\cr B'&\xi' I}\right)=
\left(\matrix{(\lambda+\lambda')I\hfill&A+A'\hfill\cr
B+B'&(\xi+\xi')I\hfill}\right).
\end{equation}

Let's define the multiplication law for the abstract matrices in the
following way:
$$\u*\u'=\left(\matrix{\lambda I&A\cr B&\xi I}\right)*
\left(\matrix{\lambda' I&A'\cr B'&\xi' I}\right)=$$
\begin{equation}\label{Dab}
=\left(\matrix{(\lambda\lambda'+\frac12\tr(AB'))I\hfill&\lambda
A'+\xi' A+\frac i2[B,B']\hfill\cr\lambda'B+\xi B'-\frac
i2[A,A']&(\xi\xi'+\frac12\tr(BA'))I\hfill}\right).
\end{equation}

It is easy to ensure, that with multiplication law defined
(\ref{Dab}), the matrices $\Sigma^i,i=1,2,\dots,7$ belong to
octonionic algebra.

Hermitian conjugation on abstract matrix algebra is introduced as
follows:
\begin{equation}\label{om1}
\left(\matrix{\lambda I&A\cr B&\xi I}\right)^+=
\left(\matrix{\lambda^*I&B^+\cr A^+&\xi^*I}\right),
\end{equation}
where $\lambda^*,\xi^*$ are complex conjugated functions and
$A^+,B^+$ are Hermitian conjugated matrices. From the definition of
Hermitian conjugation it follows that the matrices introduced above
$\Sigma^k,k=0,1,\dots,7$ are Hermitian, and matrices
$\tilde\Sigma^k,k=1,2,\dots,7$ are anti-hermitian.

Obviously, we do not leave abstract matrices space after applying
addition as well as multiplication commands introduced above to
abstract matrices.

\section{State vector on octonionic algebra}
Because of unusual multiplication law on octonionic  algebra it is
impossible to define the result of multiplication matrix and row as
we had it with standard "smart" multiplication law. That is why we
shall use not vectors-rows but matrices $\Psi(x)$ from abstract
matrix algebra, specified above:
\begin{equation}\label{opsi}
\Psi(x)=\frac1{\sqrt2}\left(\matrix{\lambda(x)I&A(x)\cr
B(x)&\xi(x)I}\right).
\end{equation}

Let's call the square of the norm of state vector
 $\Psi(x)$ as the number
$$||\Psi(x)||^2=\frac12\cdot\tr(\left(\matrix{\lambda^*I&B^+\cr
A^+&\xi^*I}\right)*\left(\matrix{\lambda I&A\cr B&\xi I}\right))=$$
\begin{equation}\label{norma}
=\lambda^*\lambda+\xi^*\xi+ \frac12\tr(B^+B+A^+A).
\end{equation}

Thus, on the algebra of extended matrix representations of
octonions, the complex scalar fields octet could be introduced.
$\varphi^i_1(x)$: $\lambda(x),\xi(x)$, $\varphi^i_1(x)$,
$\varphi^i_2(x),i=1,2,3,j=1,2,\dots,6$:
$$
\varphi_1(x)=\frac1{\sqrt2}\left(\matrix{\lambda(x)I&0\cr0&0}\right),\quad
\varphi_2(x)=\frac1{\sqrt2}\left(\matrix{0&0\cr0&\xi(x)I}\right),\quad$$
\begin{equation}\label{om1}
\varphi_{2+i}(x)=\frac1{\sqrt2}\left(\matrix{0&\varphi^i_{(1)}(x)\sigma^i\cr0&0}\right),\quad
\varphi_{5+i}(x)=\frac1{\sqrt2}\left(\matrix{0&0\cr\varphi^i_{(2)}(x)\sigma^i&0}\right)
\end{equation}
(there is no addition on $i$ index) or their linear combinations.

Further, we shall consider the potential energy on state vectors
\begin{equation}\label{potx}
V(\Psi)=-m^2\buu\Psi*\Psi+\frac f4(\buu\Psi*\Psi)^2,
\end{equation}
where $m$ and $f$ are some constants.

By the infinitesimality of state vector $\Psi(x)$ (denote the vector
as $\delta\Psi(x)$), we mean the infinitesimality of its norm, i.e.
if (\ref{norma}) then all matrix elements $\delta\Psi(x)$ are close
to zero. In this occasion by potential energy variation we mean the
following value:
$$\delta V(\Psi)=-m^2\delta\buu\Psi*\Psi
-m^2\buu\Psi*\delta\Psi+\frac f2\delta\buu\Psi*|\varphi|^2\Psi +
\frac f2\buu\Psi*|\Psi|^2\delta\Psi=$$
\begin{equation}\label{varpotx}
=\delta\buu\Psi*(-m^2\Psi+\frac f2|\Psi|^2\Psi)+ (-m^2\Psi+\frac
f2|\Psi|^2\Psi)\delta\Psi.
\end{equation}

By condition partial derivative of function  $\Psi(x)$ and
$\buu\Psi(x)$ we mean the variations (\ref{varpotx}).

The smallest value of potential energy $V(\Psi)$ is indicated when
\begin{equation}\label{potx2}
\frac{\partial V}{\partial\Psi}=0\qquad \frac{\partial
V}{\partial\buu\Psi}=0,
\end{equation}
which with $m,f>0$, as is seen from (\ref{varpotx}), gives a stable
equilibrium
$$|\Psi_0|^2=\frac{2m^2}f.$$

Let's introduce Lagrangian of the field $\Psi(x)$, the latter is
considered as Higgs field in the approach being developed.
\begin{equation}\label{gagx}
\L=\tr(\partial_\mu\buu\Psi*\partial^\mu\Psi+m^2\buu\Psi*\Psi-\frac
f4(\buu\Psi*\Psi)^2).
\end{equation}

Consider what happens with the field $\Psi(x)$ near the minimum of
potential energy as following
\begin{equation}\label{higs}
\Psi(x)=\frac1{2\sqrt2}(\frac{2m}{\sqrt{f}}+\sigma(x)+\Theta^k(x)\cdot
i\Sigma^k)\left(\matrix{0&i\sigma^3\cr0&I}\right),
\end{equation}
under supposition, that $\sigma(x)$ and $\Theta^k(x)$ are some real
functions (here and further the repeating indexes $k=1,2,\dots,7$
are implied to sum up, if other is not mentioned), so it is easy to
ensure
$$||\left(\matrix{0&i\sigma^3\cr0&I}\right)||^2=4.$$

\section{Generalization of the electro-weak interactions Lagrangian
to octoninic algebra} In Minkovskian space ($M_4$), where the
interval $ds^2$ is defined as
\begin{equation}\label{M4}
ds^2=dt^2-dx^2-dy^2-dz^2=dx_\mu dx^\mu=dt^2-dl^2=\eta_{\mu\nu}dx^\mu
dx^\nu
\end{equation}
($\mu,\nu=0,1,2,3$, here summing up on the indexes of a different
height is implied with metric tensor of Minkovsky space
$\eta_{\mu\nu}$) free Dirac equation for a spinor fields $\psi(x)$
and $\overline{\psi}(x)$ with mass of $m$ looks like \cite{Shveb}:
\begin{equation}\label{Dirsv}
(i\gamma^\mu\overrightarrow\partial_\mu-m)\psi(x)=0\qquad
\overline{\psi}(x)(i\overleftarrow\partial_\mu\gamma^\mu+m)=0,
\end{equation}
where $\overrightarrow\partial_\mu\psi(x)=\partial\psi/\partial
x^\mu=\psi_{,\mu}$. Upper and lower indexes of tensors sink and
raise respectively in $M_4$ by mutually inverse tensors
$\eta_{\mu\nu}$ and $\eta^{\mu\nu}$.
$\overline{\psi}(x)=\psi(x)^+\gamma^0,x\in M_4$,
$\gamma^\mu,\mu=0,1,2,3$ are Dirac matrices:
\begin{equation}\label{matDir}
\gamma^0=\left(\matrix{I&0\cr0&-I}\right),\qquad
\gamma^i=\left(\matrix{0&\sigma^i\cr-\sigma^i&0}\right), \quad
i=1,2,3.
\end{equation}

In the contemporary theory the interaction of a particle with an
extern field is believed to be in minimum way, which means the
derivative merely lengthens because of interacting field member
\begin{equation}\label{udl}
\partial_\mu\rightarrow\partial_\mu+
i{\bf A}_\mu
\end{equation}

Here the field variable ${\bf A}_\mu$ is introduced. Modern
perspective to field interactions is based on the fact that the
interactions comply the intern theory symmetries, which are
visualized in the field structure ${\bf A}_\mu$ provided the
interaction is introduced as it is done above. It is known from the
Maxwell electro-magnetic theory that it is Abel symmetry that
corresponds with electro-magnetic field. In work \cite{Utiyama} it
is shown the technique of introducing field variables to satisfy the
given symmetry. In such way, the incorporating of electro-magnetic
field $A_\mu$ for particles with charge ``$-e$" with minimum
connection
\begin{equation}\label{uns}
{\bf A}_\mu=-eA_\mu\cdot I,\quad I=1
\end{equation}
leads to Dirac equation in an extern electro-magnetic field
$$(i\gamma^\mu(\overrightarrow\partial_\mu-ieA_\mu)-m)\psi(x)=0$$
\begin{equation}\label{Direl}
\overline{\psi}(x)
(i(\overleftarrow\partial_\mu-ieA_\mu)\gamma^\mu+m)=0
\end{equation}

(Here the structure unit $I$ is introduced, which in our matter is
merely equal to 1.)

The union of electro-magnetic and weak interaction through the
minimum connection in the standard Weinberg-Salam theory on group
$SU(2)\times U(1)$ with fields $A_\mu^{k(1)}=A_\mu^k,k=1,2,3$ and
$A_\mu^{(2)}=B_\mu$ and charges $g^k=g$ -- $SU(2)$ group constant --
symmetry and $g^{(0)}$ -- Abel symmetry constant:
\begin{equation}\label{uns}
{\bf A}_\mu={\bf A}_\mu^{(1)}+{\bf
A}_\mu^{(2)}=gA_\mu^{k(1)}T^{k(1)}+g^{(0)}A_\mu^{(2)}T^{(2)}=\frac
i2g\sigma^kA_\mu^k+\frac i2g^{(0)}YB_\mu,
\end{equation}
$$T^{k(1)}=\frac{\sigma^k}2,\quad T^{(2)}=Y,$$
therefore the Dirac equation in extern field for fermions $\psi(x)$,
with hypercharge $Y$ (in particular in case of leptons doublet
$Y=-I$, and in case of singlet $Y=-2$) looks like \cite{Nelipa}
$$(i\gamma^\mu(\overrightarrow\partial_\mu+\frac i2g^k\sigma^kA_\mu^k
+\frac i2g^{(0)}YB_\mu)-m)\psi(x)=0$$
\begin{equation}\label{Dirsl}
\overline{\psi}(x)(i(\overleftarrow\partial_\mu-\frac
i2g^k\sigma^kA_\mu^k-\frac i2g^{(0)}YB_\mu)+m)=0
\end{equation}

The Lagrangian of free electro-weak field looks like \cite{Okun}:
\begin{equation}\label{svslo}
\L=-\frac12\tr\,G_{\mu\nu}G^{\mu\nu} -\frac14F_{\mu\nu}F^{\mu\nu}
\end{equation}

There, the first member of Lagrangian ($k=1,2,3$)
$$-\frac12\tr G_{\mu\nu}G^{\mu\nu}=-\frac12\tr((\partial_\mu
A_\nu-\partial_\nu A_\mu-ig(A_\mu A_\nu-A_\nu A_\mu))\cdot$$
\begin{equation}\label{svsl}
\cdot(\partial^\mu A^{\nu}-\partial^\nu A^{\mu}-ig(A^\nu A^\mu-A^\mu
A^\nu))),\quad A_\mu=A_\mu^k\sigma^k/2
\end{equation}
responds for non-Abel symmetry, while the second member
\begin{equation}\label{svel}
-\frac14F_{\mu\nu}F^{\mu\nu}=-\frac14(\partial_\mu
B_\nu-\partial_\nu B_\mu)(\partial^\mu B^{\nu}-\partial^\nu B^{\mu})
\end{equation}
responds for Abel one.

The spinor equation $\psi(x)$ (\ref{Dirsl}) forms to (\ref{Direl})
in case of singlet particle condition (electron) when substituted
$Y=-2$ and $g^0=e,g^k=0$. It is known from weak interactions theory
the relation between charges $g^k$ $SU(2)\times U(1)$ charges model
and charge $e$ from Dirac theory \cite{Okun}.

In the current work the effort to abstract from common way of
introducing interacting field as the field, corresponding with
certain symmetry, and it is suggested to introduce the field as
characteristic of algebraic structure of the interaction. In this
approach the existence of a symmetry, e.g. $SU(2),SU(3)$ is a
corollary of field algebraic structure. Thus, we agree the
generalization of Dirac equation to interacting field is a
generalization to octoneon fields and looks like:
$$(i\gamma^\mu(\overrightarrow\partial_\mu+\frac
i2q^aA_\mu^a\Sigma^a)-m) \Psi(x)=0$$
\begin{equation}\label{Dirvz}
\overline{\Psi}(x)(i (\overleftarrow\partial_\mu-\frac
i2q^aA_\mu^a\Sigma^a)\gamma^\mu+m)=0,
\end{equation}
where $q_a$ is a charge of the spinor $\Psi$, defined in
(\ref{opsi}), which interacts with the field $A_\mu^a(x)$,
$\Sigma^a, a=0,1,\dots,7$ are Hermitian generatrixes of octonionic
algebra.

(Note, that in (\ref{Dirvz}) matrices $\Sigma^a,a=0,1,2\dots,7$ act
to spinor intern index $\Psi(x)$, therefore it cannot be multiplied
to Dirac matrices using standard matrix multiplication.)

Consider the doublet of left-polarized leptons $L(x)$ (we shall
restrict the class by using only electron sector, consisted of
electron $e(x)=e^-(x)$ and electron neutrino $\nu=\nu_e(x)$),
singlet of right-polarized electron $e_R=e^-_R(x)$ and octet of
scalar mesons $\varphi^a(x)$ (\ref{higs}):
$$L=\left(\matrix{(\alpha_1\nu+\alpha_2e)I&A_1\nu(x)+A_2e(x)\cr
B_1\nu(x)+B_2e(x)&(\beta_1\nu+\beta_2e)I}\right)_L=\Psi_L,$$
\begin{equation}\label{matDir}
R(x)=e^-_R(x), \quad \varphi^a(x),i=0,1,\dots,7,
\end{equation}
where
$$\frac12(1+\gamma^5)\Psi=L,\qquad\frac12(1-\gamma^5)\Psi=\Psi_R,$$
\begin{equation}\label{lpm}
\frac12(1-\gamma^5)e_R=R\qquad\gamma^5=i\gamma^0\gamma^1\gamma^2\gamma^3.
\end{equation}

Matrices $A_1,A_2$ and $B_1,B_2$ look like
$A_i=\alpha_0I+\sum_1^3\alpha^k_i\sigma^k$ and
$B_i=\beta_0I+\sum_1^3\beta^k_i\sigma^k$ with numeric constants
still unknown $\alpha_i^k$ and $\beta_i^k,i=1,2, k=0,1,2,3.$ Numbers
$\alpha_{1,2}$ and $\beta_{1,2}$ are still unknown too.

Let's define the Lagrangian, localized on octoneon algebra, as
follows ($a,a'=0,1,\dots,7$)
$$\L_{oct.}=\tr(-\frac1{16}F_{\mu\nu}^{(a)}*F^{\mu\nu(a)}+$$
$$+(\partial_\mu\bux\Psi\varphi-\frac
i2q^aA_\mu^{a}\bux\Psi\varphi*\Sigma^{a})
*(\partial^\mu\Psi_\varphi+\frac
i2q^{a'}A^{\mu(a')} \Sigma^{a'}*\Psi)+$$
$$+\frac i2\overline L*\gamma_\mu(\overrightarrow\partial^\mu
L+\frac i2q^aA^{\mu(a)}\Sigma^a*L)-\frac i2\overline
L*\gamma_\mu(\overleftarrow\partial^\mu L-\frac
i2q^aA^{\mu(a)}\Sigma^{a}*L)+$$
$$+\frac i2\overline R\gamma_\mu(\overrightarrow\partial^\mu R+
iq^0A^{\mu0}R)-\frac i2\overline
R\gamma_\mu(\overleftarrow\partial^\mu R-iq^0A^{\mu0}R)-$$
\begin{equation}\label{vst}
-\tilde h\overline L*\Psi_\varphi R-\tilde h\overline
R\bux\Psi\varphi*L+m^2\bux\Psi\varphi*\Psi_\varphi-\frac
f4(\bux\Psi\varphi*\Psi_\varphi)^2)
\end{equation}

We shall use the mechanism of random symmetry violation. Exclude
Higgs' bosons ($\sigma(x)=0$) out of consideration and choose the
calibration $\Theta^k(x)=0,k=1,2,\dots,7$, which leads to
\begin{equation}\label{nkal}
\Psi_\varphi(x)=\Psi_0=\frac
m{\sqrt{2f}}\left(\matrix{0&i\sigma^3\cr0&I}\right),
\end{equation}
so we come to ($a=0,1,\dots,7$):
$$\L_{oct.}=\tr(-\frac1{16}F_{\mu\nu}^{(a)}F^{\mu\nu(a)}
+\frac14q_aq_{a'}A_\mu^{a}A^{\mu(a')}\buu\Psi_0*\Sigma^a*\Sigma^{a'}
*\Psi_0+$$
$$+\frac i2(\overline L*\gamma_\mu\partial^\mu
L-\partial^\mu\overline L\gamma_\mu*L)+\frac i2(\overline
R*\partial^\mu\gamma_\mu R-\partial^\mu\overline R\gamma_\mu*R)-$$
$$-q^aA^{\mu a}\overline L*\gamma_\mu\Sigma^a*L-
q^0A^{\mu(0)}\overline R\gamma_\mu R-$$
\begin{equation}\label{vst}
-h\overline L*\Psi_0R-h\overline R\buu\Psi_0L)+\frac {m^4}f
\end{equation}

In the standard Weinberg-Salam $SU(2)\times U(1)$ theory of weak
interaction after a random violation of symmetry, chosen
corresponding calibration, come to Lagrangian
\begin{equation}\label{glt}
\L_{ST}=-\frac14\tr(F_{\mu\nu}^{(a)}F^{\mu\nu(a)})+
\frac{g^2m^2}{2f}A_\mu^{a}A^{\mu(a)}
-\frac{gg^{(1)}m^2}fA_\mu^3B^\mu-
\end{equation}
$$+\frac{g^{(1)}}2\overline\nu_L\gamma_\mu B^\mu\nu_L+
\frac{g^{(1)}}2\overline e_L\gamma_\mu B^\mu e_L+\frac g2\overline
e_L\gamma_\mu A^{\mu3}e_L-\frac g2\overline\nu_L\gamma_\mu
A^{\mu3}\nu_L+$$
$$-\frac g2\overline\nu_L\gamma_\mu e_L(A^{\mu1}-iA^{\mu2})-
\frac g2\overline e_L\gamma_\mu\nu_L(A^{\mu1}+iA^{\mu2})-$$
$$+\frac i2(\overline e_L\gamma_\mu\partial^\mu e_L-\partial^\mu\overline
e_L\gamma_\mu e_L)+\frac
i2(\overline\nu_L\gamma_\mu\partial^\mu\nu_L-
\partial^\mu\overline\nu_L\gamma_\mu\nu_L)+$$
$$+\frac i2(\overline e_R\gamma_\mu\partial^\mu
e_R-\partial^\mu\overline e_R\gamma_\mu e_R)+g^{(1)}\overline
e_R\gamma_\mu B^\mu e_R+\frac{\sqrt2hm}{\sqrt f}(\overline
e_Le_R+\overline e_Re_L)+\frac {m^4}f$$

It is known about this Lagrangian that it diagonalized by fields
$A_\mu^a(x)$, provided charged $W,\buu W$ and neutral $Z^0$ bosons
and electromagnetical vector-potential $A_\mu$ are introduced:
\begin{equation}\label{zb}
A_\mu^3=Z^0_\mu\cos\theta+A_\mu\sin\theta,\qquad
B_\mu=A_\mu\cos\theta-Z^0_\mu\sin\theta,
\end{equation}
\begin{equation}\label{wb}
A_\mu^1=\frac1{\sqrt2}(W_\mu\cos\theta+\buu W_\mu\sin\theta),\qquad
A_\mu^2=\frac i{\sqrt2}(W_\mu\cos\theta-\buu W_\mu\sin\theta)
\end{equation}

We shall show now the solution for posed problem exists by
presenting partial solution.

Equality
$$\frac i2(\overline L*\gamma_\mu\partial^\mu
L-\partial^\mu\overline L\gamma_\mu*L)+\frac i2(\overline
R*\gamma_\mu\partial^\mu R-\partial^\mu\overline R\gamma_\mu*R)=$$
$$=\frac i2(\overline e_L\gamma_\mu\partial^\mu e_L-\partial^\mu\overline
e_L\gamma_\mu e_L)+\frac
i2(\overline\nu_L\gamma_\mu\partial^\mu\nu_L-
\partial^\mu\overline\nu_L\gamma_\mu\nu_L)+$$
$$+\frac i2(\overline e_R\gamma_\mu\partial^\mu
e_R-\partial^\mu\overline e_R\gamma_\mu e_R)$$ is provided by
constraint $R=e_R(x)$ and
$$L=\frac{c_0}{\sqrt2}\left(\matrix{0\qquad(2i\sigma^1-2\sigma^2)\nu(x)
+(y_0I+\frac{2i}3\sigma^1+\frac23\sigma^2+i\sigma^3)e(x)\cr(-\frac
i8\sigma^1+\frac18\sigma^2)\nu(x)+(-\frac{3i}8\sigma^1-\frac38\sigma^2+
\frac{9i}{16}\sigma^3)e(x)\qquad0}\right)_L$$
\begin{equation}\label{stmt}
\overline L=L^+\gamma^0,
\end{equation}
$$\overline L=\frac{c_0}{\sqrt2}\left(\matrix{0\qquad(\frac
i8\sigma^1+\frac18\sigma^2)\overline
\nu(x)+(\frac{3i}8\sigma^1-\frac38\sigma^2-
\frac{9i}{16}\sigma^3)\overline
e(x)\cr(-2i\sigma^1-2\sigma^2)\overline \nu(x)
+(y_0I-\frac{2i}3\sigma^1+\frac23\sigma^2-i\sigma^3)\overline
e(x)\qquad0}\right)_L$$

It is easy to ensure then
\begin{equation}\label{stmt}
\tr\overline L*\Sigma^1*L=c^2_0(\overline\nu_L e_L+ \overline
e_L\nu_L)$$
$$\tr\overline L*\Sigma^2*L=c^2_0(-i\overline\nu_L e_L+
i\overline e_L\nu_L)$$
$$\tr\overline L*\Sigma^3*L=c^2_0(\overline\nu_L\nu_L-
\overline e_Le_L)$$
$$\tr\overline L*L=c^2_0(\frac{257}{32}\overline\nu_L\nu_L+
(y_0^2+\frac{5729}{2304})\overline e_L e_L)
\end{equation}
Notice, that the order of multiplication does not matter when
valuating the trace (\ref{stmt}). Given $y_0^2=\frac{257}{32}-
\frac{5729}{2304}$ and $c_0^2=\frac{32}{257}$, we come to
$$\tr\overline
L*L=\overline\nu_L\nu_L+\overline e_L e_L$$

Equality
$$q^{(0)}(A^{\mu 0}\overline L\gamma_\mu*L
+A^{\mu0}\overline R\gamma_\mu R)=$$
$$=\frac{g^{(1)}}2(\overline\nu_L\gamma_\mu B^\mu\nu_L+\overline
e_L\gamma_\mu B^\mu e_L+\overline e_R\gamma_\mu B^\mu e_R)$$ is is
achieved provided $q^{(0)}c_0^2=-g^{(1)}$ and $A^{\mu0}=B^\mu$.

Equality
$$\tr(\overline L*(q^1A^{\mu1}\Sigma^1+q^2A^{\mu2}\Sigma^2+
q^3A^{\mu3}\Sigma^3)*\gamma_\mu L)=$$
$$=\frac g2\overline\nu_L\gamma_\mu e_L(A^{\mu1}-iA^{\mu2})+
\frac g2\overline e_L\gamma_\mu\nu_L(A^{\mu1}+iA^{\mu2})-$$
$$-\frac g2\overline
e_L\gamma_\mu A^{\mu3}e_L+\frac g2\overline\nu_L\gamma_\mu
A^{\mu3}\nu_L$$ is guaranteed when $q^{(k)}c_0^2=g,A^{\mu k}=A^{\mu
k},k=1,2,3$.

Equality
\begin{equation}\label{vst0}
\tilde h\overline L*\Psi_0R+\tilde h\overline
R\buu\Psi_0L=\frac{\sqrt2hm}{\sqrt f}(\overline e_Le_R+\overline
e_Re_L)
\end{equation}
is attained if $\tilde hc_0/\sqrt2=h$.

Thus, the Lagrangian is offered, given on generalized algebra of
Cally octaves, and which in particular case derives to the
Lagrangian of a standard theory of electro-weak interactions, with
condition function looking like:
$$L=\frac{c_0}{\sqrt2}\left(\matrix{0\qquad(2i\sigma^1-2\sigma^2)\nu(x)
+(y_0I+\frac{2i}3\sigma^1+\frac23\sigma^2+i\sigma^3)e(x)\cr(-\frac
i8\sigma^1+\frac18\sigma^2)\nu(x)+(-\frac{3i}8\sigma^1-\frac38\sigma^2+
\frac{9i}{16}\sigma^3)e(x)\qquad0}\right)_L=$$
$$=\frac{c_0}{\sqrt2}(\left(\matrix{0\qquad 2i\sigma^1-2\sigma^2
\cr-\frac i8\sigma^1+\frac18\sigma^2\qquad0}\right)\nu_L
+\left(\matrix{0\qquad
y_0I+\frac{2i}3\sigma^1+\frac23\sigma^2+i\sigma^3
\cr-\frac{3i}8\sigma^1-\frac38\sigma^2+
\frac{9i}{16}\sigma^3\qquad0}\right)e_L)$$

Notice, the offered Lagrangian on generalized octonionic algebra
does not differ from the Lagrangian in (\ref{glt}), given
(\ref{glt}) $a=0,\dots,7$ on chosen condition function (\ref{vst}).

\section{Entire Lagrangian of theory}
Thus, consider the integrated Lagrangian after symmetry violation,
that looks like (before fields diagonalization)
\begin{equation}\label{vst2}
\L_{oct.}=\tr(-\frac1{16}F_{\mu\nu}^{(a)}F^{\mu\nu(a)}
+\frac14q^aq^{a'}A_\mu^{a}A^{\mu(a')}\buu\Psi_0*\Sigma^a*\Sigma^{a'}
*\Psi_0+$$
$$+\frac i2(\overline L*\gamma_\mu\partial^\mu
L-\partial^\mu\overline L\gamma_\mu*L)+\frac i2(\overline
R*\partial^\mu\gamma_\mu R-\partial^\mu\overline R\gamma_\mu*R)-$$
$$-q^aA^{\mu a}\overline L*\gamma_\mu\Sigma^a*L-
q^0A^{\mu(0)}\overline R\gamma_\mu R-$$
$$-h\overline L*\Psi_0R-h\overline R\buu\Psi_0L)+\frac {m^4}f
\end{equation}

The Lagrangian, along with fields corresponding to electro-weak
interaction, has four additional fields $A^{(k)},k=4,\dots,7$, which
correspond to non-associative parts of Lagrangian. The Lagrangian is
considered on non-associative octonionic algebra, therefore it is
non-associative. The Lagrangian can be decomposed to two
expressions, associative and non-associative, according to the rule:
$$A=a_1a_2a_3=\frac12((a_1a_2)a_3+a_1(a_2a_3))+
\frac12((a_1a_2)a_3-a_1(a_2a_3))=A_{as.}+A_{nas.}$$

Three sigma matrices trace equals zero, therefore associative and
non-associative parts of four matrices trace we shall define as
follows:
\begin{equation}\label{ass4}
\tr\{\Sigma^a,\Sigma^b,\Sigma^c,\Sigma^d\}=\tr(\Sigma^a\{\Sigma^b,
\Sigma^c,\Sigma^d\}-\{\Sigma^a,\Sigma^b,\Sigma^c\}\Sigma^d)=
8\varepsilon^{abcd}.
\end{equation}

The non-associative part (\ref{vst2}) looks like:
\begin{equation}\label{maincl}
(\frac1{16}\tr F_{\mu\nu}^{(a)}F^{\mu\nu(a)})_{nas.}=
\eta^{\lambda\delta}\eta^{\mu\nu}
\varepsilon^{abcd}A_\lambda^aA_\mu^bA_\delta^cA_\nu^d.$$
$$(q^{(5)}A^{\mu(5)}\overline L*\gamma_\mu\Sigma^5*L)_{nas.}=
g_5A^{\mu(5)}(\frac{5i}4\overline\nu\gamma_\mu e-\frac{5i}4\overline
e_L\gamma_\mu\nu_L)$$
$$(q^{(6)}A^{\mu(6)}\overline L*\gamma_\mu\Sigma^6*L)_{nas.}=
g_6A^{\mu(6)}(\frac32\overline e_L\gamma_\mu
e_L+\frac54\overline\nu\gamma_\mu e+\frac54\overline
e_L\gamma_\mu\nu_L)$$ $$g_7A^{\mu(7)}\overline
L*\gamma_\mu\Sigma^7*L=0
\end{equation}

At the same time, associative part of the last three expressions
equals zero. The field $A_\mu^4$ has only associative part:
\begin{equation}\label{p4}
(q^{(4)}A^{\mu(4)}\overline L*\gamma_\mu\Sigma^4*L)_{nas.}=
g_4A^{\mu(4)}(\kappa_1\overline\nu\gamma_\mu\nu-\kappa_2\overline
e_L\gamma_\mu e_L)
\end{equation}

(Here normalization constants $\kappa_1$ and $\kappa_2$ are
approximately equal to eight and seven respectively and the
designation is introduced $q^{(k)}c_0^2=g_k,k=4,\dots,7$.)

After summing up those expressions we come to non-associative part
of the entire Lagrangian:
$$\L_{nas.}=-\eta^{\lambda\delta}\eta^{\mu\nu}
\varepsilon^{abcd}A_\lambda^aA_\mu^bA_\delta^cA_\nu^d-
\frac32g_6A^{\mu(6)}\overline e_L\gamma_\mu e_L+$$
\begin{equation}\label{nonas2}
-\frac54(g_6A^{\mu(6)}+ig_5A^{\mu(5)})\overline\nu_L\gamma_\mu
e_L-\frac54(g_6A^{\mu(6)}-ig_5A^{\mu(5)})\overline e_L\gamma_\mu
\nu_L.
\end{equation}

Therefore the entire Lagrangian on non-associative algebra consists
of parts and looks like:
\begin{equation}\label{plna}
\L_{oct.}=\L_{as.}+\L_{nas.}=
-\frac14\tr(F_{\mu\nu}^{(a)}F^{\mu\nu(a)})+
\frac{g^2m^2}{2f}A_\mu^{a}A^{\mu(a)}
-\frac{gg^{(1)}m^2}fA_\mu^3B^\mu-$$
$$+\frac{g^{(1)}}2\overline\nu_L\gamma_\mu B^\mu\nu_L+
\frac{g^{(1)}}2\overline e_L\gamma_\mu B^\mu e_L+\frac g2\overline
e_L\gamma_\mu A^{\mu3}e_L-\frac g2\overline\nu_L\gamma_\mu
A^{\mu3}\nu_L+$$
$$-\frac g2\overline\nu_L\gamma_\mu e_L(A^{\mu1}-iA^{\mu2})-
\frac g2\overline e_L\gamma_\mu\nu_L(A^{\mu1}+iA^{\mu2})-$$
$$+\frac i2(\overline e_L\gamma_\mu\partial^\mu e_L-\partial^\mu\overline
e_L\gamma_\mu e_L)+\frac
i2(\overline\nu_L\gamma_\mu\partial^\mu\nu_L-
\partial^\mu\overline\nu_L\gamma_\mu\nu_L)+$$
$$+\frac i2(\overline e_R\gamma_\mu\partial^\mu
e_R-\partial^\mu\overline e_R\gamma_\mu e_R)+g^{(1)}\overline
e_R\gamma_\mu B^\mu e_R+\frac{\sqrt2hm}{\sqrt f}(\overline
e_Le_R+\overline e_Re_L)+;$$
$$+\frac{m^4}f-g_4A^{\mu(4)}(\kappa_1\overline\nu\gamma_\mu\nu-
\kappa_2\overline e_L\gamma_\mu e_L)-$$
$$-\eta^{\lambda\delta}\eta^{\mu\nu}q^aq^bq^cq^d
\varepsilon^{abcd}A_\lambda^aA_\mu^bA_\delta^cA_\nu^d-
\frac32g_6A^{\mu(6)}\overline e_L\gamma_\mu e_L-$$
$$-\frac54(g_6A^{\mu(6)}+ig_5A^{\mu(5)})\overline\nu_L\gamma_\mu
e_L-\frac54(g_6A^{\mu(6)}-ig_5A^{\mu(5)})\overline e_L\gamma_\mu
\nu_L.
\end{equation}

The entire Lagrangian presents here with associative and
non-associative parts separated.

It seems the obtained expressions reminds the expression in the
standard model, $SU(2)$, provided new negative-charged $D$ (it is
vice versa when speaking of standard model concerning $W$-bosons)
and positive-charged $\buu D$-bosons with one charge constant $g_a$
are introduced: $g_aA=g_6A^{\mu(6)}+ig_5A^{\mu(5)}$ and $g_a\buu
A=g_6A^{\mu(6)}-ig_5A^{\mu(5)}$, but it is easy to see from
(\ref{nonas2}) such representation faces contravention second member
of equation (\ref{nonas2}). In fact, the proposed approach to define
condition in unusual way, using matrix, allows to solve the problem.
It is because when deriving the expression
$\frac32g_6A^{\mu(6)})\overline e_L\gamma_\mu e_L$ in intermediate
stage of valuation the cross members existed, which were
successfully reduced on the next stages and therefore we can rewrite
\begin{equation}\label{db}
2g_6A^{\mu(6)})\overline e_L\gamma_\mu e_L=$$
$$=(g_6A^{\mu(6)}+ig_5A^{\mu(5)})\overline e_L\gamma_\mu
e_L+(g_6A^{\mu(6)}-ig_5A^{\mu(5)})\overline e_L\gamma_\mu e_L
\end{equation}

After such modifications the non-associative Lagrangian part renews
to
\begin{equation}\label{ld}
\L_{nas.}=-\eta^{\lambda\delta}\eta^{\mu\nu}q^aq^bq^cq^d
\varepsilon^{abcd}A_\lambda^aA_\mu^bA_\delta^cA_\nu^d-
\frac34g_a(D^\mu+\buu D{\!}^\mu)\overline e_L\gamma_\mu e_L-$$
$$-\frac54g_aD^\mu\overline\nu_L\gamma_\mu
e_L-\frac54g_a\buu D{\!}^\mu\overline e_L\gamma_\mu \nu_L.
\end{equation}

Therefore the generalized Lagrangian has additional free part:
\begin{equation}\label{l0d}
\L_{0add}=-\frac12(\partial_\mu\buu D{\!}^\nu-\partial_\nu\buu
D{\!}^\mu)(\partial^\mu\buu D{\!}_\nu-\partial^\nu\buu D{\!}_\mu)+
\frac{g_a^2m^2}f\buu D{\!}^\mu D_\mu-$$ $$-\frac14(\partial_\mu
C^\nu-\partial_\nu C^\mu)(\partial^\mu C_\nu-\partial^\nu C_\mu)+
\frac{g_4^2m^2}{2f}C^\mu C_\mu-$$ $$-\frac14(\partial_\mu
E^\nu-\partial_\nu E^\mu)(\partial^\mu E_\nu-\partial^\nu E_\mu)+
\frac{g_7^2m^2}{2f}E^\mu E_\mu-$$
$$-\frac34g_a(D^\mu+\buu D{\!}^\mu)\overline e_L\gamma_\mu e_L
-\frac54g_aD^\mu\overline\nu_L\gamma_\mu e_L-\frac54g_a\buu
D{\!}^\mu\overline e_L\gamma_\mu \nu_L.
\end{equation}

As it follows from the latter, it is natural to expect omitting two
oppositely charged bosons simultaneously, imitating neutral current.
(It is noteworthy, there is no resembling neutral neutrino current
in the Lagrangian (\ref{ld}).)

\section{Conclusion}
The result of the Lagrangian introduced is the appearance of new
neutral massive $C-$current and one new massive charged $D-$current.
Unfortunately, the proposed theory, in the frameset of a model in
the Minkowskian space, does not offer any exact values of new bosons
masses.

The important issue is the appearance of a massive neutral particle,
$E-$boson, which does not interact with matter, but interact with
$A_\mu^{(a)}$ fields by means of non-associative connection.

\section{Acknowledgements}
The author expresses his gratitude to the participants of the
Friedmann Seminar for Theoretical Physics (St. Petersburg) for
profound discussions.

\end{document}